# Nanoscale domains in strained epitaxial BiFeO$_3$ thin Films on LaSrAlO$_4$ Substrate

Zuhuang Chen, Lu You, Chuanwei Huang, Yajun Qi, Junling Wang, Thirumany Sritharan [a] and Lang Chen [b]

School of Materials Science and Engineering, 50 Nanyang Avenue,

Nanyang Technological University, Singapore 639798, Singapore

**Abstract**

BiFeO$_3$ thin films with various thicknesses were grown epitaxially on (001) LaSrAlO$_4$ single crystal substrates using pulsed laser deposition. High resolution x-ray diffraction measurements revealed that a tetragonal-like phase with $c$-lattice constant ~4.65 Å is stabilized by a large misfit strain. Besides, a rhombohedral-like phase with $c$-lattice constant ~3.99 Å was also detected at film thickness of ~50 nm and above to relieve large misfit strains. In-plane piezoelectric force microscopy studies showed clear signals and self-assembled nanoscale stripe domain structure for the tetragonal-like regions. These findings suggest a complex picture of nanoscale domain patterns in BiFeO$_3$ thin films subjected to large compressive strains.

____________________________________________

Electronic mail:

a) ASSRITHARAN@ntu.edu.sg

b) langchen@ntu.edu.sg

Multiferroic materials have attracted considerable interest recently because of the intriguing fundamental physics and wide range of potential applications.[1,2] Among them, BiFeO$_3$ (BFO) is the only known single phase multiferroic material which possesses simultaneous ferroelectric and magnetic orders at room temperature.[3,4] Multiferroic properties, robust ferroelectricity and lead-free nature make BFO a prime candidate for next-generation devices including non-volatile memories and lead-free piezoelectric.[5,6]

Epitaxial strain resulting from the difference in lattice parameter between the film and the substrate, can often strongly influence the structure and properties of the films.[7] At room temperature, bulk BFO exhibits a rhombohedrally distorted perovskite structure, with space group $R3c$ and lattice parameters $a_r = 3.96$ Å and $\alpha_r = 89.4°$ .[8] Previous studies have reported that the remnant polarization,[9] the domain structure,[10] the domain wall stability[11] and even the phase symmetry[12] in BFO films can be altered by epitaxial strain. One of the most remarkable findings is the prediction of a metastable phase of strained BFO thin film with P4mm symmetry ($a$=3.67 Å and $c$=4.65 Å) by first-principles calculations, which does not exist in bulk but can be stabilized by large compressive strain.[13-15] Recent experimental results confirm the existence of a metastable phase in BFO films grown on LaAlO$_3$ (LAO) substrates, and suggest that this is not an exact tetragonal P4mm but is monoclinic with $Cc$ symmetry (tetragonal-like phase).[16,17] This phenomenon is of great interest because this metastable phase is predicted to have a giant polarization value of about 150 μC/cm$^2$. Its highly strained structure where the tetragonal-like (T) and rhombohedral-like (R) phases coexist, exhibits huge electromechanical responses.[15,17] However, there are still some discrepancies regarding the domain structures and properties of BFO films grown on LAO substrates.[16,17]

Furthermore, the LAO substrates are usually heavily twinned, which may further complicate the study of BFO crystal structure and may even deteriorate the physical properties of epitaxial films.[18]

In this letter, we report the crystal and domain structure of epitaxial highly strained BFO thin films grown on (001) LaSrAlO$_4$ (LSAO) single crystal substrates. At room temperature, LSAO has the structure of K$_2$NiF$_4$ (I4/mmm space group) with $a = b = 3.755$ Å and $c = 12.6$ Å.[19] One main advantage of using LSAO substrate is the absence of twinning and structural phase transitions in the processing temperature window.[19] Therefore, LSAO substrates have better crystalline quality and less mosaic patterns than LAO substrates do, as demonstrated by the x-ray diffraction (XRD) rocking curves shown in Fig. 1(a). Furthermore, LSAO has a larger lattice misfit of 5.2% with BFO as compared to a 4.3% misfit of LAO. Therefore, LSAO is expected to be more effective than LAO in stabilizing the T phase of BFO. Consequently, the critical thickness where a strain-induced transition[17] appears is expected to be larger in films on LSAO than on LAO. A larger critical thickness would avoid the leakage problem common in ultrathin films and facilitate electrical property measurements.

BFO thin films with several thicknesses ranging from 10 nm to 210 nm were grown on (001) LSAO single crystal substrates by pulsed laser deposition with a KrF excimer laser ($\lambda = 248$ nm).[20] The deposition temperature and the oxygen pressure were 700 °C and 100 mTorr, respectively. After deposition, the samples were cooled to room temperature at a rate of 5°C/min in 1 atm of oxygen. The structure and crystalline quality of the BFO thin films were studied by a high-resolution X-ray diffractometer (Panalytical

X-pert Pro). Piezoelectric force microscopy (PFM) investigations were carried out on an Asylum Research MFP-3D atomic force microscope using Pt/Ir-coated tips.

Figure 1(b) shows XRD $\theta-2\theta$ scans of BFO films with different thicknesses grown on LSAO substrates. Only the 00$l$ type diffraction peaks were observed. For the 20 nm film, only the diffraction peaks from the T phase were detected with no second phase or impurities. The out-of-plane $c$ lattice parameter calculated from the position of the 001 peak is $c$= 4.65 Å, which is much larger than the bulk value of 3.96 Å but is close to that of T-phase predicted by first principles calculations.[13,15] The result demonstrates that T-phase can be stabilized by large misfit strain. When the film thickness increased to 80 nm, weak peaks corresponding to the 00$l$ reflections of the R-phase ($c$~3.99 Å) appeared, implying the coexistence of T and R phases at this thickness. With a further increase in film thickness, the intensity of the 001 peak of R phase steadily increased relatively to those of T phase. This indicates an increasing R phase content with an increase of film thickness. Fig. 1 (c) shows the XRD data across the 001 peak of the T phase for different thicknesses. For film thicknesses up to ~50nm, the 001 peak position remains unchanged, which indicates that the BFO films are fully strained. For higher thicknesses there is a systematic shift in the 001 peak position of the T phase towards lower diffraction angles. The corresponding $c$-lattice parameter finally approaches the value of 4.675 Å, as can be seen in the plot of $c$-lattice versus thickness in Fig. 1(d). Because the in-plane lattice parameter of T-phase is smaller than that of LSAO, therefore, T-phase is under a tensile strain instead of a compressive strain, according to Poisson's relationship, out-of-plane lattice is compressed for ultrathin films. Out-of-plane c-lattice will relax gradually with the further increase of film thickness by the formation of R-phase, and finally relaxes to value close to the "strain-free" metastable T-phase.[17] The full width at half maximum (FWHM) determined from the rocking curves of 001 peaks of T-phase range from 0.06º to 0.14º for all samples, implying that all the BFO films have high crystallinity.

The XRD results above suggest that ~50nm is the critical thickness below which the BFO films are fully strained and have single T-phase structure. For higher thicknesses the lattice strain is progressively relaxed by the formation of the R-phase. The phase evolution of the epitaxial BFO films can also be verified by the change of topography with the increase of thickness. Fig. 2 shows characteristic atomic force microscopy (AFM) topography images of the samples, indicating a strong thickness-dependence of the film morphology. The topography images consist of two distinct features. The plateau feature arises from the T phase while the stripe-like contrast arises from the multiphase T and R areas. The latter contrast is attributed to the difference in $c$-lattice parameters between two phases. As seen evident in Fig 2, the stripe area is hardly observed at the thickness of 20 nm. The multiphase stripe area is clear at higher thicknesses and its fraction increases with the increase of film thickness, which is consistent with the XRD results. Further, the stripes are in groups of two orientations of which are approximately perpendicular to each other, and nearly parallel to the substrate edges. Such stripes were reported previously in BFO films on LAO[17] but they occupied a larger area than in our samples. This demonstrates that LSAO, due to the smaller in-plane lattice parameter can stabilize the T phase more effectively at the same film thickness which indirectly supports our claim that the T-phase could be stable up to larger film thicknesses when LSAO substrate is used.

The polarization in bulk BFO is oriented along <111>, which gives rise to eight possible domain variants.[21] The domain structure and related properties of BFO thin films grown on low mismatch substrates (SrTiO$_3$ and DyScO$_3$) are well studied.[22] However, the domain structure in films grown on substrates that could give rise to large compressive mismatch strains have not been fully understood to date. Figs. 3 (a) and (b) show two examples of the domain structure of 55 nm and 110 nm thick BFO films respectively grown on LSAO. The surface of the strained BFO thin film is atomically smooth. The 55-nm thick film shows families of small-sized stripes in Fig. 3a (i) while large-sized ones appear in Fig. 3b (i) for the 110-nm thick film. Additionally, the area fraction of striped regions in the 55-nm film is also smaller than that in the 110-nm one. Out-of-plane phase image shows uniform contrast in Figs. 3a (ii) and 3b (ii), suggesting that all out-of-plane polarizations are pointing in one direction. It is interesting to note from the out-of-plane amplitude images shown in Figs. 3a (iii) and 3b (iii) that the piezoelectric responses in stripe areas are larger than those of plateau areas, which further supports our proposition that the stripe areas are two-phase mixtures with large $d_{33}$ values. Contrary to the reports of Bea et al.[16] and Zeches et al.[17], it is found that the in-plane PFM images shown in Figs. 3a (iv) and 3b (iv), have regular in-plane domain contrasts. The in-plane domain structure exhibited here is similar to those striped domains in BFO films on SrTiO$_3$ substrate but with a smaller domain width.[23] Moreover, the stripe domains here are oriented in the [110] direction, rather than [100] reported for BFO films on SrTiO$_3$,[22] which indicates that the polarization vector probably lies in the (100) planes and tilts at certain angles with respect to [001] direction.

The distinct in-plane PFM contrast further suggests that the T phase in this highly distorted BFO films is not perfectly tetragonal with (001) polarization, but perhaps is monoclinic with the polarization vector at an angle to the *c*-axis, and therefore it gives some detectable in-plane projections.[15] The averages of stripe domain width *w* are plotted in Fig.4 in a log-log scale as a function of film thickness *d*. A least squares fit of the domain period *w* as a function of film thickness *d* yields a power law $w \propto d^{\gamma}$, with a scaling factor $\gamma$ ~0.51 (shown in Fig. 4), which agrees well with the typical values of 0.5 for all ferroic domains in thin films.[24]

In summary, we have shown the tetragonal-like $BiFeO_3$ phase can be stabilized on $LaSrAlO_4$ substrates with detectable in-plane polarizations and observed self-assembled nanoscale stripe domains with a scaling factor 0.51, in agreement with typical ferroic domain's scaling. The complex nanoscale domain patterns in $BiFeO_3$ thin films under large compressive strains provide rich varieties for further studies of domain and domain walls, magnetism and even photovoltaic properties and need to be carefully studied.

**Acknowledgement**

The authors acknowledge the valuable discussions with Prof. Ramesh and Dr. Eddie Chu and the supports from Nangyang Technological University and Ministry of Education of Singapore under Projects No. AcRF RG 21/07 and No. ARC 16/08.

**Figure Captions**

FIG. 1. (Color online) (a) XRD rocking curves around the (001) peak of LAO and (004) peak of LSAO substrate. (b) XRD θ-2θ scans of BFO films with different thicknesses grown on LSAO substrates. (c) Detailed θ-2θ scans around (001) peaks of BFO. (d) Out-of-plane lattice parameter as a function of film thickness.

FIG. 2. (Color online) AFM images of (a) 20 nm, (b) 55 nm, (c) 60 nm, (d) 110 nm, (e) 140 nm, and (f) 210 nm thick BFO films on LSAO substrates illustrating the thickness-dependence of the film's morphology. All images are 5×5 μm$^2$ and all AFM images have a height scale of 8 nm.

FIG. 3. (Color online) (i) Topography, (ii) Out-of-plane phase, (iii) out of plane amplitude and (iv) in-plane PFM images of (a) 55nm and (b) 110nm thick BFO films on LSAO substrates. All images are 5×5 μm$^2$ and all AFM images have a height scale of 8 nm.

FIG. 4. (Color online) A scaling plotting between in-plane domain width and film thickness is shown. The straight line is a least-square fitting yields a scaling exponent of 0.51. The inset pictures show some examples of in-plane PFM images of BFO films with different film thicknesses.

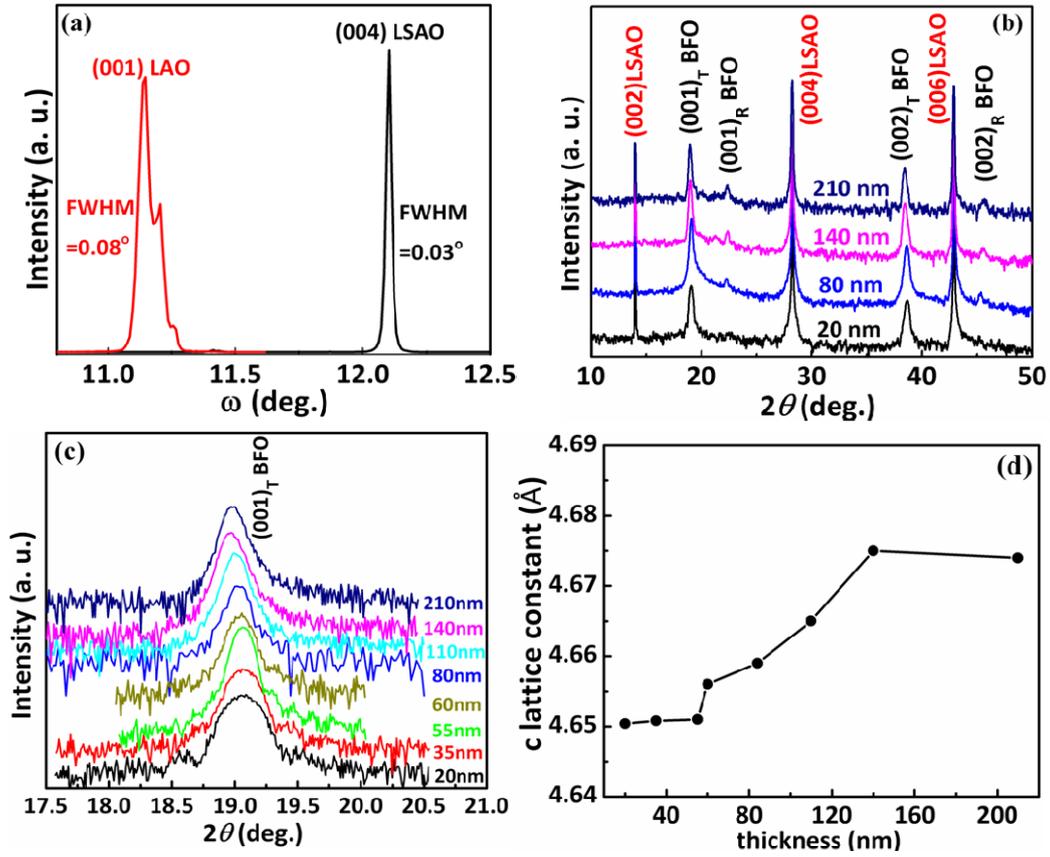

FIG. 1. (Color online) (a) XRD rocking curves around the (001) peak of LAO and (004) peak of LSAO substrate. (b) XRD θ-2θ scans of BFO films with different thicknesses grown on LSAO substrates. (c) Detailed θ-2θ scans around (001) peaks of BFO. (d) Out-of-plane lattice parameter as a function of film thickness.

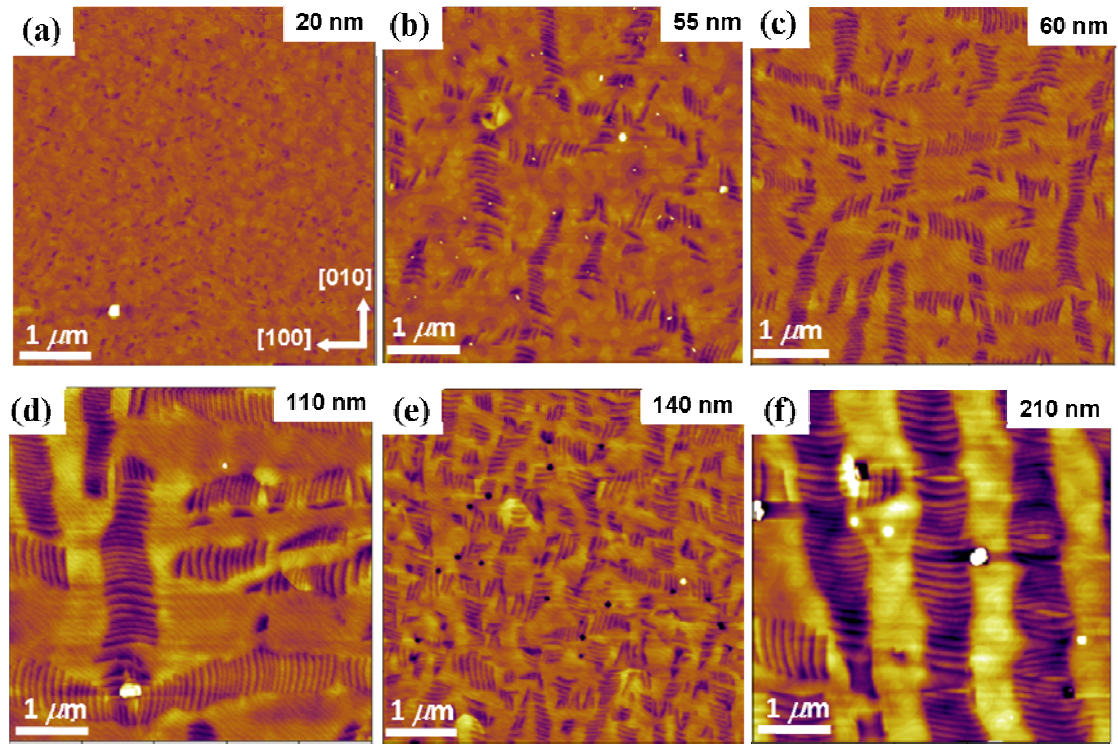

FIG. 2. (Color online) AFM images of (a) 20 nm, (b) 55 nm, (c) 60 nm, (d) 110 nm, (e) 140 nm, and (f) 210 nm thick BFO films on LSAO substrates illustrating the thickness-dependence of the film's morphology. All images are 5×5 μm$^2$ and all AFM images have a height scale of 8 nm.

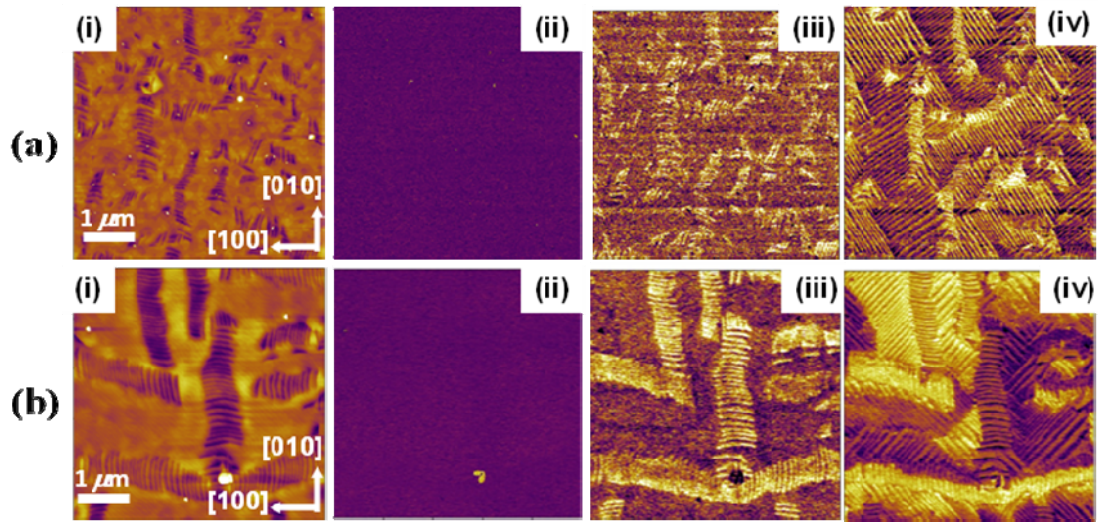

FIG. 3. (Color online) (i) Topography, (ii) Out-of-plane phase, (iii) out of plane amplitude and (iv) in-plane PFM images of (a) 55nm and (b) 110nm thick BFO films on LSAO substrates. All images are 5×5 μm$^2$ and all AFM images have a height scale of 8 nm.

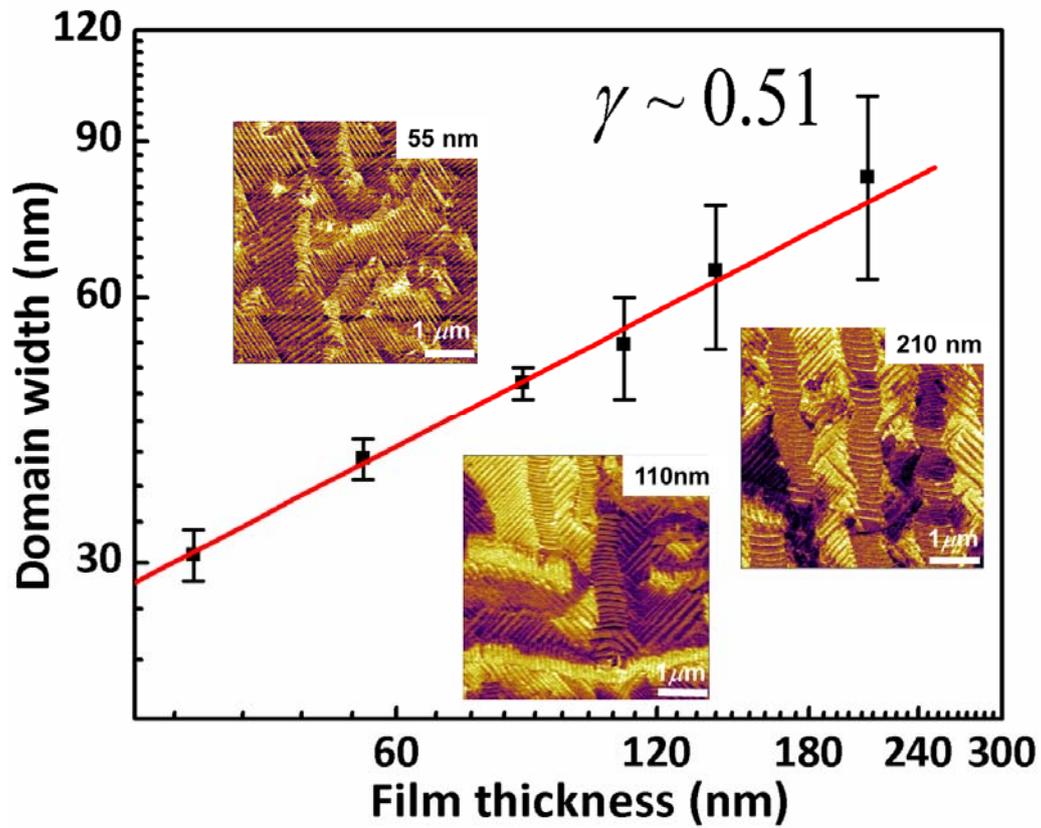

FIG. 4. (Color online) A scaling plotting between in-plane domain width and film thickness is shown. The straight line is a least-square fitting yields a scaling exponent of 0.51. The inset pictures show some examples of in-plane PFM images of BFO films with different film thicknesses.